\documentclass[12pt]{article}
\usepackage{amsfonts}

\usepackage{graphicx}
\usepackage{amsmath}

\begin{document}

\begin{titlepage}
\begin{center}
 \today \hfill UFR-HEP 01/01\\
\vskip 2cm {\LARGE {\bf D-branes \\ ~\\ \bf on
Noncommutative Orbifolds}}
 \vskip 2 cm
 {\large E. M. Sahraoui\footnote{lphe@fsr.ac.ma}, E.
H. Saidi\footnote{H-saidi@fsr.ac.ma} }\\
 \vskip 0.5cm $^{1,2}$ UFR-High Energy Physics, Department of
Physics, Faculty of Sciences,\\ Rabat University, Av. Ibn Battouta
B.P. 1014, Morocco.\\
\bigskip
 and
\\
\vskip 0.5cm $^1$ Dipartimento di Fisica Teorica dell'Universit\`a
di Torono\\ Via Pietro Giuria 1. Torino 10125, Italy.

\vskip 1cm
{\bf  Abstract}
\end{center}
We study tachyon condensation on noncommutative toric orbifolds
with a $\mathbb{Z}_{2}$
 discrete group and explore the various kinds of brane bound states
arising in the case of irrational values of the $B$-field. We show
that $\mathbb{Z}_{2}$ symmetry of the orbifolds incorporates
naturally anti-branes in the spectrum and leads to equivalent
results as those obtained by starting from an original pair of
$D$-$\overline{D}$  system on quantum torii. A specific analysis
is deserved to the irrational representation of NC orbifolds and
to the unstable bound states generated by the condensation.
\end{titlepage}
\newpage

\section{Introduction}

Recently a great interest has been given to the study of tachyon
condensation using non-commutative (NC) geometry \cite{a}.
 Starting from non-BPS $D$-branes of string theory and turning
on an antisymmetric NS-NS $B$-field, one gets a condensation of
tachyon fields on the world volume of the lower dimensional branes
\cite{b}. The key step in the derivation of this result is based
on the computation of the vacuum energy configurations after
neglecting the kinetic part of the string field effective action
in front of the potential term
 \cite{c}. The energy of the soliton is shown to
be proportional to the trace of projectors on the ground states of the non
commutative algebra Hilbert space \cite{d}.

Using GMS construction \cite{e}, NC tachyon condensation has been first\
studied in \cite{f} and especially electric fluxons obtained from original non-BPS $D$%
-branes on the non compact Moyal spaces in the presence of a constant $B$-field
and second in \cite{g}, see also \cite{h,i}, in terms of $D0$-branes
by starting from a $D2$-brane on a NC two torus. In \cite{j}, we have
extended the above mentioned results to more general cases; in particular to
higher dimensional torii where we have shown the existence of general
unstable bound states decaying into $D$0-$D$2 ones and suggested an effective
potential to describe these brane states. In the study performed in \cite{j}%
, it has remained however a discussion regarding the
$D$-$\overline{D}$ systems on non-commutative compact manifolds.
Such systems are interesting in the study of unstable $D$ brane
systems in type II string theory as they go beyond the theory of
non BPS $D$-branes recovered by restricting to couplings invariant
under $(-)^{F_L}$ \cite{k} and were already considered from the
view point of complex tachyon condensation on the quantum two
torus. But here we shall reconsider this analysis differently
using the discrete symmetries of the NC brane world volumes.\ The
present study may be then viewed as a continuation of the analysis
we initiated in \cite{j} by first completing our previous results
to the case of non-BPS branes on non-commutative toric orbifolds.
We also study  the $D0$-$D2$ bound states of \cite{g} and explore
the
nature of branes generated by the tachyon condensation in NC $\mathbb{T}%
_{\theta }^{2}/\mathbb{Z}_{2}$. Then we analyze the spectrum of
brane bound states one gets generally when studying tachyon
condensation of $D$-branes wrapped on higher dimensional
orbifolds.

The aim of this work is to start from an original non-BPS $D2l$-brane and
study the NC solitons in $\mathbb{Z}_{2}$ toric orbifolds for both rational
and irrational NC $\theta _{i}^{\prime }$s parameters.
 We first study tachyon in $\mathbb{T}_{\theta }^{2}/\mathbb{Z}%
_{2}$ and  consider the
extension to higher dimensional compact orbifolds. Then, we derive the
various kinds of solitons and explore all types of bound states one gets after
the condensation. Finally we study the $D$-$\overline{D}$ brane systems
from different views; once by starting from a pair of $D$-$\overline{%
D}$ brane on quantum torii and second by considering only a $D$-brane on
orbifolds.

The paper is organized as follows. In section 2, we describe two
different representations of the NC $\mathbb{Z}_{2\text{ }}$
orbifolds according to whether the NC parameter $\theta $\ is
rational or not. In section 3,
 we analyze the two cases of non BPS branes on the NC orbifold torus
 constructed in section 2 depending on whether $\theta$'s are rational
 or not. In section 4, we study the $D$-$\overline{D}$ brane systems
either by using standard analysis based on starting from an original pair
of
$D$-$\overline{D}$ branes or by using the $\mathbb{Z}_{2}$ symmetry of the
orbifolds. The last section is devoted for discussions and conclusion.

After this study had been completed several works treating in the
same line appeared \cite{l,m,n,o}.
\section{Solitons in non-commutative Orbifolds}

Here we study briefly the construction of non-commutative orbifolds
$\mathbb{T%
}_{\mathbf{\theta }}^{2l}/\mathbb{Z}_{2}$ by starting from the usual
realisation of non-commutative $\mathbb{T}_{\mathbf{\theta }}^{2l}$ and
imposing $\mathbb{Z}_{2}$ symmetry. Then we study its rational and
irrational representations and give the projectors on the ground state configurations
which will be used later in the building of solitons in $\mathbb{T}_{%
\mathbf{\theta }}^{2l}/\mathbb{Z}_{2}$.\\
 To start recall that the $\mathbb{T}%
_{\mathbf{\theta }}^{2l}$ we will be considering is roughly speaking given
by the product of $l$ non-commutative two dimensional torii $\mathbb{T}%
_{\theta _{i}}^{2}$ generated by a system of $l$ unitary operator pairs $%
\left( U_{i},V_{i}\right) $ satisfying the algebra \cite{j}
\begin{eqnarray}
U_{i}V_{i} &=&\text{e}^{-i2\pi \theta _{i}}V_{i}U_{i},\qquad i=1,...,l  \label{nctorus} \\
U_{i}V_{j} &=&V_{j}U_{i};\quad i\neq j,  \label{nctorus2}
\end{eqnarray}

for which we will shall refer from now on as $\frak{A}_{\theta }$, the non-commutative
algebra of functionals on the torus $\mathbb{T}_{\mathbf{\theta }}^{2l}$. Note that because of eq(%
\ref{nctorus},\ref{nctorus2}), this non-commutative algebra $\frak{A}_{\theta }$ may be also
defined as the tensor product of $l$\ factors $\frak{A}_{\theta _{i}}$ as $%
\frak{A}_{\theta }=\otimes _{i=1}^{l}\frak{A}_{\theta _{i}}$; where each $%
\frak{A}_{\theta _{i}}$ factor is associated with the non-commutative torus $%
\mathbb{T}_{\theta _{i}}^{2}$; and the corresponding $\left(
U_{i},V_{i}\right) $ pairs are realized as the exponential of the
non-commutative coordinates $\left( x^{2i-1},x^{2i}\right) $ of $\mathbb{T}%
_{\theta _{i}}^{2}$. For later use we prefer to denote the
coordinates pairs of the\ $\mathbb{T}_{\theta _{i}}^{2}$\
non-commutative torii by the capital letters $\left(
X^{2i-1},X^{2i}\right) $ while those of the commutative ones by
small Latin letters. Thus we have
\begin{eqnarray}
U_{i} &=&\text{e}^{\frac{i2\pi }{R_{2i-1}}X^{2i-1}};\quad  \label{2orbtorus}
\\
V_{i} &=&\text{e}^{\frac{i2\pi }{R_{2i}}X^{2i}},\quad i=1,\ldots ,l
\end{eqnarray}
where the $R_{j}$'s are the one cycles radii of the $2l$ dimensional torus.\
Note also that the adjoint operator pairs $\left( U_{i}^{+},V_{i}^{+}\right)
$, which may read as
\begin{eqnarray}
U_{i}^{+} &=&e^{-\frac{i2\pi }{R_{2i-1}}X^{2i-1}};\quad \\
V_{i}^{+} &=&e^{-\frac{i2\pi }{R_{2i}}X^{2i}},\quad i=1,\ldots ,l,  \notag
\end{eqnarray}
satisfy the following equivalent identities.
\begin{eqnarray}
U_{i}^{+}V_{i}^{+} &=&\text{e}^{-i2\pi \theta _{i}}V_{i}^{+}U_{i}^{+}, \\
U_{i}^{+}V_{j}^{+} &=&V_{j}^{+}U_{i}^{+};\quad i\neq j.
\end{eqnarray}
Now we turn to define the orbifold $\mathbb{T}_{\mathbf{\theta }}^{2l}/%
\mathbb{Z}_{2}$ as the quotient of $\mathbb{T}_{\mathbf{\theta }}^{2l}$ by $%
\mathbb{Z}_{2}$. In fact, an element $\Omega $\ of
$\mathbb{Z}_{2}$\ \ identifies the local coordinates $\left\{
x^{I}\right\} $ $\left( x^{I}\equiv x^{I}+2\pi R^{I}\right) $ with
$\left\{ -x^{I}\right\} $, $\left( 1\leq I\leq 2l\right) $. While
the $\left( U_{i},V_{i}\right) $ and $\left(
U_{i}^{+},V_{i}^{+}\right) $
pairs defining the non-commutative
orbifold are related as:
\begin{eqnarray}
\Omega U_{i}\Omega &=&U_{i}^{+},\quad \Omega ^{2}=I_{id}  \notag \\
\Omega V_{i}\Omega &=&V_{i}^{+},\quad 1\leq i\leq l.  \label{orbicons}
\end{eqnarray}

To solve these eqs,\ we will distinguish two cases according to whether the $%
\theta _{i}$'s are rational or not and so two different
representations for the orbifold. We shall use both of these
representations; that's why we propose to first describe them briefly on the
simple case of the non-commutative $\mathbb{T}_{\theta }^{2}/\mathbb{Z}_{2}$
orbifold; then we extend our result to the general situation. To begin
  note that natural solutions of eqs(\ref
{orbicons}) are given by setting the $U_{i}$'s an $V_{i}$'s as:
\begin{equation}\label{uvmatrices}
U_{i}=\left(
\begin{array}{cc}
A_{i} & E_{i} \\
C_{i} & D_{i}
\end{array}
\right) ,\quad V_{i}=\left(
\begin{array}{cc}
B_{i} & F_{i} \\
G_{i} & H_{i}
\end{array}
\right) ,
\end{equation}
where $A_{i}$, $B_{i}$, $C_{i}$, $D_{i}$, $E_{i}$, $F_{i}$, $G_{i}$ and\ $%
H_{i}$ are operators to be determined later. Putting this
change in eqs(\ref{orbicons}), one gets:
\begin{eqnarray}
D_{i} &=&A_{i}^{+},\quad C_{i}=E_{i}^{+}, \notag\\
H_{i} &=&B_{i}^{+},\quad G_{i}=F_{i}^{+}.  \label{abcdefgh}
\end{eqnarray}
At this level we would like to emphasize that the above
constraints are
inherent to the $\mathbb{Z}_{2}$ orbifold in the sense that all $A_{i}$, $%
B_{i}$, $C_{i}$, $D_{i}$, $E_{i}$, $F_{i}$, $G_{i}$ and\ $H_{i}$
matrix operators should have the same dimension. This will
constitute a key property in distinguishing rational and
irrational $\mathbb{Z}_{2}$ orbifolds and will play an important
role when we discuss the solitons problem. Moreover, and as far as
the study of solitons on non-commutative spaces is concerned, one
can further impose $E_{i}=C_{i}=0$.\
This special choice will be justified later on when we build the projectors $%
\Pi _{i}$ and $\mathcal{P}_{N_{i}+M_{i}\theta _{i}}$; see eqs(\ref{ratpro}) and (\ref{irratpro}),
on the vacuum field configurations. For convenience we shall
also require that $F_{i}=G_{i}=0$. This representation is a particular one
which will help us to simplify the analysis; it is reducible into
irreducible factors $(A_{i},B_{i})$ and their hermitian conjugates.

\bigskip

\textit{Rational representations }

\quad This representation \ corresponds to rational values of $\theta
_{i}=q_{i}/p_{i}$, where $p_{i}$ and $q_{i}$ are\ mutually coprime integers.
The $A_{i}$ and $B_{i}$ operators are given by the following finite $%
p_{i}\times p_{i}$ matrices
\begin{equation}
A_{i}=\left[
\begin{array}{cccc}
1 & 0 & \cdots & 0 \\
0 & \omega _{i} & \ddots & 0 \\
\cdots & \cdots & \ddots & 0 \\
0 & 0 & \cdots & \omega _{i}^{p_{i}-1}
\end{array}
\right] ;\quad \quad B_{i}=\left[
\begin{array}{cccc}
0 & 1 & \cdots & 0 \\
0 & 0 & \ddots & 0 \\
\cdots & \cdots & \ddots & 1 \\
1 & 0 & \cdots & 0
\end{array}
\right]  \label{matrices}
\end{equation}
where $\omega _{i}=$e$^{i2\pi q_{i}/p_{i}}$. Note in passing that $%
A_{i}^{p_{i}}$ and $B_{i}^{p_{i}}$ act as the $p_{i}\times p_{i}$ identity
operator $I_{p_{i}\times p_{i}},$ and so is $U_{i}^{p_{i}}$ and $%
V_{i}^{p_{i}}$. Therefore any element $e_{i}$ of the non-commutative algebra
$\frak{A}_{\theta _{i}}$ associated to $\mathbb{T}_{\theta }^{2}/\mathbb{Z}%
_{2}$ has a finite expansion
\begin{equation}
e_{i}=\sum_{n,m=0}^{p_{i}-1}(e_{i})_{nm}U_{i}^{n}V_{i}^{m}.
\end{equation}

In the matrix representation presented above, the $U_{i}$ generators are
given by diagonal matrices; a feature which allows to build the usual $k_{i}$%
-th rank $\Pi _{i}$ projectors $(\Pi
_{i})_{k_{i}}=diag(1,1,...,1,0,...,0;1,1,...,1,0,...,0),$ as a
series of the $U_{i}$'s as shown here below:
\begin{equation}\label{ratpro}
(\Pi _{i})_{k_i}=\sum_{n=0}^{p_{i}-1}(e_{i})_{n0}U_{i}^{n}.
\end{equation}

A direct check shows that the $%
(e_{i})_{n0}$ coefficients are given by $(e_{i})_{n0}=\frac{1}{p_{i}}\frac{%
1-\omega _{i}^{-nk_i}}{1-\omega _{i}^{-n}}$. Note that according to eq(\ref{uvmatrices}),
the $\Pi _{i}$ projectors may be rewritten in terms of the $A_{i}^{\prime }$%
s and their hermitian $A_{i}^{+}\equiv \overline{A}_{i}$\ conjugates as
\begin{equation}
(\Pi _{i})_{k_{i}}=\sum_{n=0}^{p_{i}-1}(e_{i})_{n0}\left(
\begin{array}{cc}
(A_{i})^{n} & 0 \\
0 & (\overline{A}_{i})^{n}
\end{array}
\right) ,  \label{ratpro}
\end{equation}
or equivalently
\begin{equation*}
\Pi _{i}=\pi _{i}\oplus \overline{\pi }_{i},
\end{equation*}
where
\begin{eqnarray}
(\pi _{i})_{k_{i}} &=&\sum_{n=0}^{p_{i}-1}(e_{i})_{n0}A_{i}^{n}, \\
(\overline{\pi }_{i})_{k_{i}} &=&\sum_{n=0}^{p_{i}-1}(\overline{e}%
_{i})_{(p_{i}-n)0}\overline{A}_{i}^{n}.
\end{eqnarray}
Since the trace on $\frak{A}_{\theta _{i}}$\ is given by Tr$(\Pi
_{i})_{k_{i}}=(e_{i})_{00}+(\overline{e}_{i})_{p_{i}0}=(k_i+{\overline
k}_i)=2k_{i}$, then the range of $k_{i}$\ which is bounded as
$0\leq k_{i}\leq p_{i}$ will be interpreted as\ giving the number
of $D0$ and $\overline{D0}$ branes  one obtains from the study of
\ the condensation of a non BPS $D2$-brane on the NC
$\mathbb{T}_{\theta }^{2}/\mathbb{Z}_{2}$. Note moreover that the
$\Pi _{i}$ projector we built satisfy naturally
\begin{equation}\label{identification}
\Omega \Pi _{i}\Omega =\Pi _{i},\qquad \Omega ^{2}=I_{id}.
\end{equation}

This identification requires that the number of $D0$-branes should be equal
to the number of $\overline{D}0$-branes. If one relaxes this condition, one
may also considers the building of field configurations\  where
the numbers of $D0$-branes and $\overline{D}0$-branes are not necessary
equal. This violates however the $\mathbb{Z}_{2}$ symmetry
\begin{equation}\label{hermitic}
D_{i}=A_{i}^{+},\qquad H_{i}=B_{i}^{+},
\end{equation}
and so will be omitted.

\bigskip

\textit{Irrational representations}

\quad The generalization of the previous case to irrational
$\theta _{i}$'s is not automatic first because it uses more
involved functional analysis and second its interpretation in
terms of systems of \ $D0$-$\overline{D}0$ branes bound states.
Following the same lines as for the rational case by working in a
representation in which $U_{i}$ is diagonal and using results on
the non-commutative irrational torus, one can take the following
realization
\begin{eqnarray}
\langle x_{2i-1}^{\prime }|U_{i}|x_{2i-1}\rangle &=&\left[
\begin{array}{cc}
\text{e}^{ix_{2i-1}} & 0 \\
0 & \text{e}^{-ix_{2i-1}}
\end{array}
\right] \delta \left( x_{2i-1}-x_{2i-1}^{\prime }\right) , \notag\\
\langle x_{2i-1}^{\prime }|V_{i}|x_{2i-1}\rangle &=&\left[
\begin{array}{cc}
\delta \left( x_{2i-1}+\theta _{i}-x_{2i-1}^{\prime }\right) & 0 \\
0 & \delta \left( x_{2i-1}-\theta _{i}-x_{2i-1}^{\prime }\right)
\end{array}
\right] ,\label{irratrep}
\end{eqnarray}
where we have set $R_{I}=\frac{1}{2\pi }$ for commodity. The $U_{i}$ and\ $%
V_{i}$\ operators depend on the $x^{2i-1}$ variables and not on the $x^{2i}$%
  ones. To construct the non-commutative orbifold projector operators on the
position space generated by the continuous basis vectors $\left\{
|x_{2i-1}\rangle \times |x_{2i}\rangle \right\} $, one may
consider in a first attempt functions of the diagonal operator
$U_{i}$. An apparently adequate candidate for the function
$f(U_{i})$ is given by:
\begin{equation}\label{irratpro}
\langle x_{2i-1}^{\prime }|(\mathcal{P}_{i})|x_{2i-1}\rangle =\langle
x_{2i-1}^{\prime }|f(U_{i})|x_{2i-1}\rangle =\left(
\begin{array}{cc}
\kappa _{i} & 0 \\
0 & \kappa _{i}
\end{array}
\right) \delta \left( x_{2i-1}-x_{2i-1}^{\prime }\right) ,
\end{equation}
\ \ for\ $0\leq x_{2i-1}\leq \kappa _{i}$, while for $\kappa
_{i}<x_{2i-1}\leq 1$ we have
\begin{equation*}
\langle x_{2i-1}^{\prime }|f(U_{i})|x_{2i-1}\rangle =0,
\end{equation*}
where $\kappa _{i}$ is a priori a real parameter lying between zero and one.
Though this choice of $(\mathcal{P}_{i})$\ ensures that it is hermitian, satisfies $%
(\mathcal{P}_{i})^{2}=(\mathcal{P}_{i}),$ it fails however as the trace Tr($\mathcal{P}_{i})$ is
not an integer in general since,
\begin{equation}
\text{Tr}(\mathcal{P}_{i})=\int \text{d}x_{2i-1}\langle x_{2i-1}|(\mathcal{P}
_{i})|x_{2i-1}\rangle =2\kappa _{i}.
\end{equation}
This trace is not acceptable, it contradicts the expected spectrum dictated
by the group $\mathbf{K}_{0}\left( \frak{A}^{o}_{\theta _{i}}\right) =\mathbb{Z}%
+\theta _{i}\mathbb{Z}$, as $\kappa _{i}$ is not quantized. To
overcome this difficulty one should use both the $U_{i}$ and
$V_{i}$ operators in building $\mathcal{P}_{i}$ instead of using
$U_{i}$ alone; this will allow to solve our problem and also
incorporate explicitly the non-commutativity parameter into the
game. Guided by the result on the non-commutative irrational
torii, a
class of solutions for the projector operators in agreement with $\mathbf{K}%
_{0}(\frak{A}^{o}_{\theta _{i}}$) reads as:
\begin{equation}
\mathcal{P}_{N_{i}+M_{i}\theta _{i}}=\left(
\begin{array}{cc}
P_{n_{i}+m_{i}\theta _{i}} & 0 \\
0 & \overline{P}_{\overline{n}_{1}+\overline{m}_{1}\theta _{i}}
\end{array}
\right) ,  \label{irrprojector}
\end{equation}
where $N_{i}$ and $M_{i}$ stand \ respectively for the multi-indices $%
(n_{i},m_{i})$ and $(\overline{n_{i}},\overline{m}_{i})$ and where
\begin{eqnarray}
P_{n_{i}+m_{i}\theta _{i}} &=&\left( \overline{B}_{i}^{m_{i}}\right) \left(
g\left( \overline{A}_{i}\right) \right) +f\left( A_{i}\right) +g\left(
A_{i}\right) B_{i}^{m_{i}};\notag \\
\overline{P}_{\overline{n}_{1}+\overline{m}_{1}\theta _{i}} &=&\overline{g}%
\left( A_{i}\right) B_{i}^{\overline{m_{i}}}+\overline{f}\left( \overline{A}%
_{i}\right) +\left( \overline{B}_{i}^{\overline{m_{i}}}\right) \left(
\overline{g}\left( \overline{A}_{i}\right) \right) .\label{pbarp}
\end{eqnarray}
The eigenvalues of the functions $f(A_{i}),\overline{f}\left( \overline{A}%
_{i}\right) $ and $g\left( A_{i}\right) $ and $\left({\overline g}\left( \overline{A}%
_{i}\right) \right) $ are given by:
\begin{equation}
f\left( A_{i}\right) =\left\{
\begin{array}{ll}
x^{2i-1}/\epsilon _{i} & x^{2i-1}\in \left[ 0,\epsilon _{i}\right] \\
1 & x^{2i-1}\in \left[ \epsilon _{i},\theta _{i}\right] \\
1-\left( x^{2i-1}-(n_{i}+m_{i}\theta _{i})\right) /\epsilon _{i}\quad \quad
& x^{2i-1}\in \left[ \theta _{i},\theta _{i}+\epsilon _{i}\right] \\
0 & x^{2i-1}\in \left[ \theta _{i}+\epsilon _{i},1\right]
\end{array}
\right.
\end{equation}
\begin{equation}
\overline{f}\left( \overline{A}_{i}\right) =\left\{
\begin{array}{ll}
-x^{2i-1}/\epsilon _{i} & x^{2i-1}\in \left[ -\epsilon _{i},0\right] \\
1 & x^{2i-1}\in \left[ -\theta _{i},-\epsilon _{i}\right] \\
1+\left( x^{2i-1}+({\overline n}_{i}+{\overline m}_{i}\theta _{i})\right) /\epsilon _{i}\quad \quad
& x^{2i-1}\in \left[ -\theta _{i}-\epsilon _{i},-\theta _{i}\right] \\
0 & x^{2i-1}\in \left[ -1,-\theta _{i}-\epsilon _{i}\right]
\end{array}
\right.
\end{equation}
and
\begin{eqnarray}
g\left( A_{i}\right) &=&\left\{
\begin{array}{cc}
\frac{(x^{2i-1}\left( \epsilon _{i}-x^{2i-1}\right) }{\epsilon _{i}}^{1/2} &
x^{2i-1}\in \left[ 0,\epsilon _{i}\right] \\
0 & x^{2i-1}\in \left[ \epsilon _{i},1\right]
\end{array}
\right. , \\
\left( \overline{g}\left( \overline{A}_{i}\right) \right) &=&\left\{
\begin{array}{cc}
\frac{(-x^{2i-1}\left( \epsilon _{i}+x^{2i-1}\right) }{\epsilon _{i}}^{1/2}
& x^{2i-1}\in \left[ -\epsilon _{i},0\right] \\
0 & x^{2i-1}\in \left[ -1,-\epsilon _{i}\right]
\end{array}
\right. ,
\end{eqnarray}
where $\epsilon _{i}$\ is a small parameter which physically may be
interpreted as a regulator parameter. Actually, the above result extends the
constructions considered in \cite{g} for the Powers-Rieffel projectors and
in \cite{j} for quantum torii. Observe here that the integers $(n,m)$ and
$(\overline{n},\overline{m})$ carried by the $P$ and $\overline{P}$
projectors are not necessary equal as the constraint eqs(\ref{abcdefgh}) are not
violated. This means that in the irrational realization of the $\mathbb{Z}%
_{2} $ orbifold eq (\ref{identification}) is no longer a necessary constraint contrary to the
rational case. So for irrational $\mathbb{Z}_{2}$ orbifolds, one will be
dealing with richer systems of $D-\overline{D}$ branes as we shall see later
on.

Having given the representations of $\frak{A}^{o}_{\theta _{i}}$ for $\mathbb{T}%
_{\theta }^{2}/\mathbb{Z}_{2}$, we turn now to extend them to $\mathbb{T}_{%
\mathbf{\theta }}^{2l}/\mathbb{Z}_{2}$. For fixed $l$, we have generally $%
2^{l}$ possibilities depending on whether the $\theta _{i}$'s are rational
or irrational. If all $\theta _{i}$'s are rational, i.e. $\theta
_{i}=q_{i}/p_{i}$ the $U_{i}$ and $V_{i}$ are given by similar eqs to eq (%
\ref{matrices}). If instead all $\theta _{i}$'s are irrational, the $U_{i}$%
's and $V_{i}$'s are given by
\begin{eqnarray}
\langle \mathbf{x}^{\prime }|U_{i}|\mathbf{x}\rangle &=&\left[
\begin{array}{cc}
\text{e}^{ix_{2i-1}}\delta \left( \mathbf{x-x}^{\prime }\right) & 0 \\
0 & \text{e}^{-ix_{2i-1}}\delta \left( \mathbf{x-x}^{\prime }\right)
\end{array}
\right] , \\
\langle \mathbf{x}^{\prime }|V_{i}|\mathbf{x}\rangle &=&\left[
\begin{array}{cc}
\delta ^{2l}\left( \mathbf{x}+\theta _{i}-\mathbf{x}^{\prime }\right) & 0 \\
0 & \delta ^{2l}\left( \mathbf{x}-\theta _{i}-\mathbf{x}^{\prime }\right)
\end{array}
\right] .
\end{eqnarray}

We can also have the case where part of the $\theta _{i}$'s are rational and
the others are irrational. In this case the $U_{i}$' s and $V_{i}$'s are
given by mixing the representations (\ref{matrices}) and (\ref{irratrep}).

The projectors $\mathcal{P}_{\left\{ \theta _{1},\ldots ,\theta _{l}\right\}
}$ on the position basis $\{|\mathbf{x}\rangle
=|(x_{1},x_{2},...,x_{2l-1},x_{2l})\rangle \}$ for $\mathbb{T}_{\mathbf{%
\theta }}^{2l}/\mathbb{Z}_{2}$ have then several forms depending on whether the $%
\theta _{i}$'s are rational or irrational. Denoting by $\mathcal{P}_{\left\{
\theta _{i}\right\} }$ the projector operator associated to $\theta _{i}$
which is given by either eq (\ref{matrices}) or (\ref{irrprojector}), we
have
\begin{equation}
\mathcal{P}_{\left\{ \theta _{1},\ldots ,\theta _{l}\right\} }=\overset{l}{%
\underset{i=1}{\otimes }}\mathcal{P}_{\left\{ \theta _{i}\right\} }.
\label{gtprojector}
\end{equation}

From eq (\ref{gtprojector}), one learns that there are a priori
$2^{l}$ solutions. However if one identifies operators that are
related under permutations of positions, one ends then with $l$
different objects. Note that the trace of eq (\ref{gtprojector})
is given by the product of the traces of the individual projectors
$\mathcal{P}_{\left\{\theta _{i}\right\}}$; i.e
\begin{equation}
\text{Tr}\mathcal{P}_{\left\{ \theta _{1},\ldots ,\theta
_{l}\right\} }=\prod_{i=1}^{l}\text{Tr}\mathcal{P}_{\left\{ \theta
_{i}\right\} }=\prod_{i=1}^{l}(\text{Tr}P_{\left\{ \theta
_{i}\right\} }+\text{Tr}\overline{P}_{\left\{ \theta _{i}\right\}
}). \label{traceproj}
\end{equation}

\section{Non-BPS Branes on Orbifolds}

Consider a non-BPS $D2l$-brane in presence of a $B_{\mu \nu }$
field on the NC orbifold
$\mathcal{O}$=$\mathbb{T}^{2l}_{\theta}/\mathbb{Z}_{2}$ we
introduced earlier and study the field configurations minimizing
the total energy $E(T)$ of the tachyon living on the $D2l$-brane
world volume. Keeping only the tachyon field $T(x)$ and
integrating out all other fields, the string field theory
effective action $\mathcal{S}=\mathcal{S}(T(x))$ reads as
\begin{equation}\label{sftaction}
\mathcal{S}=\frac{C_{D2l}}{G_{S}}\int_{\mathcal{O}}\text{d}^{2l}\text{x}%
\sqrt{G}\left( \frac{1}{2}f\left( \ast T\right) G^{\mu \nu }\partial _{\mu
}T\partial _{\nu }T+\cdots +V\left( \ast T\right) \right) ,
\end{equation}
where\ $G_{S}$ is the open string coupling constant and $G_{\mu \nu }$, $%
C_{D2l}$ and the factor $f\left( t\right) $ are related by
\begin{eqnarray}
G_{\mu \nu } &=&g_{\mu \nu }-\left( 2\pi \alpha ^{\prime }\right) ^{2}\left(
Bg^{-1}B\right) _{\mu \nu }, \\
C_{D2l} &=&\ G_{S}M_{D2l}.
\end{eqnarray}
$G_{\mu \nu }$ and $g_{\mu \nu }$ being the effective open string
and closed string metrics respectively and $f(T)$ is the effective
coupling normalized like $f(0)=0$ and $f(t_{\text{max}})=1$ as
suggested by Sen's conjecture.

In large non-commutativity, the kinetic term of the tachyon is neglected so
that the action (\ref{sftaction}) reduces to $\mathcal{S}=\frac{C_{D2q}}{G_{S}}\int_{\mathcal{O}}$d$^{2q}$x$\sqrt{G}%
V\left( \ast T\right) $. Therefore, the total energy $E(T)$\ reads, upon
taking $G_{\mu \nu }=\delta _{_{\mu \nu }}$, as
\begin{equation}
E(T)=M_{D2l}\text{Tr}V(T),
\end{equation}
where $M_{D2l}$ denotes the mass of the original $D2l$-brane and
the trace Tr is normalized as Tr$\mathbf{1}=1$. Extremisation of
$E(T)$\ is achieved as usual by using the GMS approach\ which
shows that the tachyon field
configuration is proportional to the\ projectors in the $\frak{A}_{\mathbf{%
\theta }}$ NC algebra.\ The idea is based on taking the tachyon
field $T(x)$ as
\begin{equation}
T(x)=\sum_{r}t_{r}\frak{p}_{r}(x),
\end{equation}
where $\frak{p}_{r}\frak{\ }$are mutually orthogonal projectors of $\mathcal{%
A}_{\theta }$ standing for $\Pi _{k}$ eq(\ref{ratpro}) or $\mathcal{P}%
_{n+m\theta }$ eq(\ref{irrprojector})\ depending on whether
$\theta $ is rational or irrational. Then using the GMS method,
the total energy of the vacuum configurations can be shown to be
proportional to the trace of the projectors as shown here below
\begin{equation}
E=M_{D2l}\sum_{i}V(t_{i})\text{Tr}(\frak{p}_{i}),  \label{energy}
\end{equation}
the $t_{i}$'s are the critical values solving $\frac{dV(t)}{dt}=0$.
Following the Sen's conjecture the tachyon potential $V(t)$ has two extrema;
one minimum at the origin $t_{\text{min}}=0$ with $V(t_{\text{min}})=0$ and
a maximum at $t_{\text{max}}$ with $V(t_{\text{max}})=1$ and so
\begin{equation}\label{tsolution}
T(x)=t_{\text{max}}\frak{p}(x).
\end{equation}

Recall in passing that $t_{\text{max}}$ and $t_{\text{min}}$ describe
respectively an unstable local maximum representing the space filling $D2$%
-brane $(V(t_{\text{max}})=1)$, and a local minimum representing the closed
string vacuum without any D-branes $(V(t_{\text{min}})=0)$. The orbifold
vacuum field configurations (\ref{energy}) lie between the original $D2l$%
-brane associated with $T=t_{\text{max}}\mathbf{1}$ $\mathbf{=}t_{\text{max}%
}\sum_{r=1}^{\infty }\frak{p}_{r}(x)$, and the complete tachyon condensation
$\left( T=0\right) $. Let us discuss these configurations.

\subsection{Branes on rational orbifolds}

To start consider the case of a $D2$-brane on the NC orbifold $%
\mathbb{T}_{\theta }^{2}/\mathbb{Z}_{2}$ described by\ the non-commutative
coordinate operators $(A,B)$ and their conjugates eqs(\ref{matrices}). Level $k$ solitons
on the rational orbifold\ $\mathbb{T}_{\theta }^{2}/\mathbb{Z}_{2}$ are
given by
\begin{equation}\label{rtsolution}
T_{k}(x)=t_{\text{max}}(\Pi )_{k}=\frac{t_{\text{max}}}{p}\sum_{n=0}^{p-1}%
\frac{1-\omega ^{-nk}}{1-\omega ^{-n}}\left(
\begin{array}{cc}
(A)^{n} & 0 \\
0 & (\overline{A})^{n}
\end{array}
\right) .
\end{equation}
Therefore the energy (\ref{energy}) reads as
\begin{equation}\label{renergy}
E=\frac{t_{\text{max}}}{2p}M_{Dq}(k+\overline{k}).
\end{equation}
where $k=\overline{k}$.\ Following the analysis of \cite{g} made
for the quantum two torus and using general results on orbifold
symmetries, in particular in compactification of Matrix model on
orbifolds \cite{k}, the above equation can be interpreted in a
nice way in terms of $D0$-$\overline{D}0$ branes systems. At the
fixed points of the orbifold eq(\ref{renergy}) describes $D0$
branes but also anti-D0 branes as a consequence of the
$\mathbb{Z}_{2}$ orbifold symmetry. More precisely it describes
$k$ coincident $D0$-branes and $k$ coincident
$\overline{D}0$-branes, which\ at low energies, are described by a
non perturbative effective field theory with a $U(k)\times U(k)$
gauge symmetry. This result is automatically extendable to the
higher dimensional orbifolds $\mathbb{T}_{\theta
}^{2l}/\mathbb{Z}_{2}$. Thus
starting from an unstable $D2l$ brane on $\mathbb{T}_{\theta }^{2l}/\mathbb{Z%
}_{2}$ and following the same lines we described earlier for the leading $%
l=1$ case, one gets:
\begin{equation}
E_{l}=\sum_{i=1}^{l}\frac{t_{\text{max}}}{2p_{i}}M_{D2l}(k_{i}+\overline{k}%
_{i}).
\end{equation}
This brane configuration has $\sum_{i=1}^{l}k_{i}$ number of $D0$-branes and
the same number of anti $D0$-branes so that the gauge group of the
underlying effective field theory is: $U(\sum_{i=1}^{l}k_{i})\times
U(\sum_{i=1}^{l}k_{i})$. Finally we should note that a part the doubling of
the spectrum, the analysis of tachyon condensation on NC
orbifolds is quite similar to the condensation on rational torii. For
details see \cite{j}.

\subsection{ Branes on irrational orbifolds}

Non-BPS $D2$-branes wrapped on the irrational orbifold
$\mathbb{T}_{\theta }^{2}/\mathbb{Z}_{2}$ lead to solitons
proportional to the projectors on the ground states of the Hilbert
space of the non-commutative algebra $\frak{A}^{o}_{\theta}$. The
$\frak{p}_{r}$ projectors, which are given by an appropriate
generalization of the Power-Rieffels operators, reads now as:
\begin{equation}\label{irtsolution}
T_{N+M\theta }=t_{\text{max}}\mathcal{P}_{N+M\theta }=t_{\text{max}}\left(
\begin{array}{cc}
P_{n+m\theta } & 0 \\
0 & \overline{P}_{\overline{n}+\overline{m}\theta }
\end{array}
\right).
\end{equation}
The total energy of this field configuration is then
\begin{equation}\label{irrenergy}
E=t_{\text{max}}M_{D2}[(n+\overline{n})+(m+\overline{m})\theta ],
\end{equation}
while its index $Ind(T)$ is
\begin{equation}\label{index}
Ind(T)=t_{\text{max}}M_{D2}[(n-\overline{n})+(m-\overline{m})\theta ],
\end{equation}

Moreover,\ setting $t_{\text{max}}=1$\ and\ using the interpretation in
terms of branes and anti branes as well as the following mass spectrum
relations
\begin{eqnarray}
M_{D2} &=&M_{\overline{D}2}=\sqrt{2}\frac{R_{1}R_{2}}{g_{s}\left( \alpha
^{\prime }\right) ^{\frac{3}{2}}}\left( \left[ 1+\left( 2\pi \alpha ^{\prime
}B\right) ^{2}\right] ^{\frac{1}{2}}\right) , \\
M_{D0} &=&M_{\overline{D}0}=\sqrt{2}\frac{1}{g_{s}\left( \alpha ^{\prime
}\right) ^{\frac{1}{2}}}=\theta M_{D2}=\theta M_{\overline{D}2}.
\end{eqnarray}
where $M_{D2}$ $(=M_{\overline{D}2})$ and $M_{D0}$ $(=M_{\overline{D}0})$
are respectively the masses of the non BPS $D2$\ $\left( \overline{D}%
2\right) $ and of $D0$ $(\overline{D}0)$ branes and where $B=\frac{1}{2\pi
R_{1}R_{2}\theta }$, the energy of the vacuum configuration can be split as
\begin{equation}
E=(nM_{D2}+\overline{n}M_{\overline{D}2})+\theta (mM_{D0}+\overline{m}M_{%
\overline{D}0}),  \label{boundenergy}
\end{equation}
This formula means that the original $M_{D2}$ annihilates to $D0$-$D2$, $D0$-%
$\overline{D}2$, $\overline{D}0$-$D2$ and $\overline{D}0$-$\overline{D}2$
bound states. This interpretation is not new as it has been first
given in \cite{g} in the context of the study of $D2$-brane on the
irrational two torus annihilating to $D0$-$D2$ bound states. It is dictated
by the analysis on tachyon condensation and also supported by $T$-duality as
well as the exact mass spectrum of the $\{mD0$, $\overline{m}\overline{D}0$,
$nD2$, $\overline{n}\overline{D}2\}$ bound states
\begin{equation}
M_{\left( n,m\right) }=M_{\left( \overline{n},\overline{m}\right) }=\sqrt{2}%
\frac{R_{1}R_{2}}{g_{s}\left( \alpha ^{\prime }\right) ^{\frac{3}{2}}}\left[
1+\left( 2\pi \alpha ^{\prime }B_{\text{eff}}\right) ^{2}\right] ^{\frac{1}{2%
}},
\end{equation}
where the effective $B_{\text{eff}}$ field is
\begin{equation}
B_{\text{eff}}=B+\frac{1}{2\pi R_{1}R_{2}}\frac{m}{n}.
\end{equation}
Taking the\ large limit of $\left( 2\pi \alpha ^{\prime }B_{\text{eff}%
}\right) $, eq(\ref{boundenergy}) is then reproduced. The novelty
here is that one is in presence of a richer spectrum of bound
states involving branes and anti branes. The remaining analysis is
like in the quantum two torus case. In what follows, let us give
some general results of branes on NC orbifolds.

\section{More on Bound states}

So far, we have discussed vacuum energy configurations of a
tachyon field on orbifolds for the two representations: rational
and irrational. In the former, we have learned that the energy of
the $D$ and $\overline{D}$ brane states is bounded by the initial
$D$-brane mass $M_{D}$ and we have
interpreted this feature by saying that the number of $D$ and $\overline{D}$%
-branes filling in the orbifold is limited; the number of $D$ or
anti $D$-brane can not exceed p, the weight of the rational
representation of the orbifold. In the irrational case, the total
energy is still bounded by $M_{D}$, however due to the continuous
property of states density, one is left with an
unusual system of $D$ and anti $D$ brane bound states. In the case of a $D2$%
-brane on a NC irrational two torus, the bound states one has are
of type $mD0$-$nD2$; they were interpreted as describing $k$
$D0$-branes where $k$ is the greatest common divisor of the $n$
and $m$ integers. This idea has been extended in \cite{j} to the
case of a $D$-branes wrapped on higher dimensional torii. There we
have found general bound states involving $\{D2s,\quad 0\le s\le
l\}$
 brane systems.
 In the case of $D2$-brane on the $\mathbb{T}_{\theta }^{2}/\mathbb{Z}_{2}$
orbifold, the situation is quite similar, the main difference with $\mathbb{T%
}_{\theta }^{2}$ is the presence of extra bound states involving
anti $D$ branes as shown in eq(\ref{boundenergy}). This result
extends naturally to higher dimensions. To determine the gauge
symmetry of the underlying field
theory, we proceed as for the NC torus since the system still inherits the torus $%
T $-duality symmetry. In deed, since the total energy spectrum bound states
configurations
\begin{equation}
E\left( \mathcal{P}_{N+M\theta }\right) =\left[ (n+\overline{n})+(m+%
\overline{m})\theta \right] \frac{V}{G_{s}}
\end{equation}
is invariant under $SL(2,\mathbb{Z}),$ i.e
\begin{equation*}
\left(
\begin{array}{c}
\alpha ^{\prime } \\
\beta ^{\prime }
\end{array}
\right) =\left(
\begin{array}{cc}
s & -r \\
-q & p
\end{array}
\right) \left(
\begin{array}{c}
\alpha \\
\beta
\end{array}
\right) .
\end{equation*}
where $sp-rq=1$, and where $\alpha ,\beta $ stand for n, m, $\overline{n}$
 and $\overline{m}$. one can usually set \ $k=n/r$ and $\overline{k}=%
\overline{n}/r$. Therefore the gauge symmetry is $U(k)\times U(\overline{k})$
with $k$\ and $\overline{k}$\ the greatest common divisor of $\left(
n,m\right) $\ and $\left( \overline{n},\overline{m}\right) $\ respectively.
In other words, the system is similar to parallel $k$\ $D0$-brane and $%
\overline{k}$\ $\overline{D}0$-branes.

\section{D-\={D} systems}

In this section we would like to give some comments regarding
brane anti brane condensations and also compare our results on
tachyons condensation on orbifolds with the analysis of \cite{g}
dealing with $D$-$\overline{D}$ systems on NC torii There, it has
been shown that starting from a pair of $D2$-brane anti-$D2$-brane
system on the two non commutative torus, the total energy may be
written in the large $B$ limit as:
\begin{equation}\label{barenergy}
E(T,\overline{T})=M_{D2}Tr[V(1-T\overline{T})+V(1-\overline{T}T)],
\end{equation}
where $T$ and $\overline{T}$ are the complex tachyon and its conjugates.
These tachyon fields are treated a little bit differently from the usual
real tachyon field; they are interpreted as as linear maps interpolating
between the two Hilbert spaces $\mathcal{H}$ and $\overline{\mathcal{H}}$\
\begin{equation*}
T:\mathcal{H}\rightarrow \overline{\mathcal{H}},\qquad \overline{T}:\overline{%
\mathcal{H}}\rightarrow \mathcal{H}
\end{equation*}
and satisfying the following eqs which may roughly be thought of
as a complexification of the GMS anzats
\begin{equation}\label{tbart}
T=T\overline{T}T,\quad \overline{T}=\overline{T}T\overline{T}.
\end{equation}
Using these relations, projectors $\Pi $ and $\overline{\Pi}$ on the vacuum subspaces
of $\mathcal{H}$ and $\overline{\mathcal{H}}$ have been built in \cite{h} and can
be shown to be given by
\begin{eqnarray}\label{pitbart}
\Pi &=&1-T\overline{T}, \\
\overline{\Pi } &=&1-\overline{T}T.
\end{eqnarray}
These are self adjoint operators verifying $\Pi ^{2}=\Pi $ and $\overline{%
\Pi }^{2}=\overline{\Pi }$ together with other relations
specifying their actions on $T$ and $\overline{T}$. Using the
modified Power-Rieffels realization of $\Pi $ and $\overline{\Pi
}$ on the non commutative torus, see eqs(34) of \cite{g}, as well
as the above identities (\ref{pitbart}), the total energy formula
(\ref{barenergy}) becomes then
\begin{eqnarray*}
E(T,\overline{T}) &=&M_{D2}{\rm Tr}[\Pi +\overline{\Pi }] \\
&=&M_{D2}[(n+m)\theta
]+M_{\overline{D}2}[\overline{n}+\overline{m}\theta ].
\end{eqnarray*}
Similar relations such as for the index $Ind(T,\overline{T})$ may
be also written down. We shall not give them explicitly here, they
may be obtained immediately from the analysis of section 3 and the
following discussion. What we want to make now is to give some
comments about this system. The first thing one may note about the
above construction concerns the potential of the complex tachyon.
This is a striking $U(1)$ invariant function from which one
already learns the essential about the vacuum
configuration projectors and so about the condensation of the $D$-$\overline{%
D}$ systems. This construction is a priori extendable to
$D$-$\overline{D}$ systems involving various numbers of original
brane and anti branes and seems to lead to more general GMS type
relations extending eqs(\ref{tbart}) and (\ref{pitbart}). For
example if instead of the complex $T$ and $\overline{T}$ fields,
one
considers the following $2\times 2$ matrix valued fields $T$ and $\overline{T}$%
\begin{eqnarray*}
T &=&T_{0}I+\sum_{i=1}^{3}T_{i}\sigma ^{i}, \\ \overline{T}
&=&\overline{T}_{0}I+\sum_{i=1}^{3}\overline{T}_{i}\sigma ^{i},
\end{eqnarray*}
where ($T_{0},T_{i})$ and ($\overline{T}_{0},\overline{T}_{i})$
are four complex tachyon fields and their complex conjugates and
where the $\sigma ^{i}$'s are the usual Pauli matrices, the
analogue of eqs(\ref{tbart}) and (\ref{pitbart}) reads as
\begin{eqnarray*}
T &=&TT^{\dagger }T,\text{ \ \ }T^{\dagger }=T^{\dagger }TT^{\dagger }, \\
\Pi &=&1-TT^{\dagger },\overline{\Pi }=1-T^{\dagger }T.
\end{eqnarray*}
In this case the energy and the index of the vacuum configuration
is invariant under a $U(2)$ automorphism group rotating the
different tachyons. Nevertheless if one is not insisting on\
automorphism symmetries, in particular on the above mentioned
$U(1)$ symmetry of \cite{g} and considers
instead real tachyons $T_{1}$ and $T_{2}$ interchanged under a $\mathbb{Z}%
_{2}$ symmetry, one gets the same spectrum of brane states in the
condensation. Indeed the total energy and the index of the real
$D2$ brane-anti $D2$ brane configuration on the non-commutative
two torus reads in large non-commutativity
\begin{eqnarray*}
E &=&E(T_{1})+E(T_{2})=M_{D2}TrV(T_{1})+M_{\overline{D2}}TrV(T_{2}), \\
I_{nd} &:=&Index(T_{1},T_{2})=M_{D2}TrV(T_{1})-M_{\overline{D2}}TrV(T_{2}).
\end{eqnarray*}
Since in this limit $T_{1}$ and $T_{2}$ are not coupled, they obey then
exactly to the standard rule giving the GMS solitons on the quantum two
torus. A direct check shows that after straightforward calculations, one
ends with the same states spectrum as in the previous complex approach and
so the same conclusion. However and as far as concluding about the $D$-$%
\overline{D}$ system is concerned, we still have something else to add. This
concerns the way the $\mathbb{Z}_{2}$ symmetry involved in the game. Besides
the two ways we discussed above where $\mathbb{Z}_{2}$ acts directly on the
tachyon fields but not on the noncommutative manifold on which the brane and
anti brane are wrapped, one can also consider the general case where $%
\mathbb{Z}_{2}$ does act on the quantum manifold. This is the case of $%
\mathbb{Z}_{2}$-orbifolds we have been discussing in this paper.
Let us briefly summarize the thing.

Through our discussion, we have learned in the case of an original
$D2$ brane wrapped on\ the non commutative orbifolds
$\mathcal{O}_{\theta }$ that due the constraint
eqs(\ref{orbicons}), $\mathbb{Z}_{2}$ symmetry leads to a natural
doubling of the torii Hilbert space spectrum interpreted as
corresponding to brane and anti branes. The latest emerge
automatically because of the orbifold Z$_{2}$ symmetry and do not
need to introduce an original anti-D2 brane. For selfcontainess of
this discussion let recall some of our results and compare to what
we said in the beginning of this section. In the case of
irrational orbifolds for instance, we have shown that the vacuum
field configuration for the tachyon field is
\begin{equation*}
T_{N+M\theta }=t_{\text{max}}\left(
\begin{array}{cc}
P_{n+m\theta } & 0 \\
0 & \overline{P}_{\overline{n}+\overline{m}\theta }
\end{array}
\right) ,
\end{equation*}
where{\small \ }$P_{n+m\theta }$ and $\overline{P}_{\overline{n}+\overline{m}%
\theta }$ are generalization of Power Rieffels projectors given by
eq(\ref{pbarp}) and the energy $E$ and the index $I_{nd}$ of the
soliton configurations are identical to (\ref{barenergy}). To see
why these results match, we give hereafter an algebraic argument\
to explain the equivalence of the approaches.

The non-commutative orbifold $\mathcal{O}_{\theta }$ associated
with the non commutative algebra $\mathcal{A}^{o}_{\theta
}(\mathcal{O}rbifold)$ \
is much more constrained than the usual non-commutative algebra \ $\mathcal{A%
}_{\theta }(Torus)$. Because of these constraints,
representations $\mathcal{H}^{o}(\mathcal{O}rbifold)$ of
  $\mathcal{A}_{\theta }%
\mathcal{(O}rbifold\mathcal{)}$ have twice the dimension of $\mathcal{H}%
(Torus)\equiv \mathcal{H};$ the $\mathcal{A}_{\theta }\mathcal{(}Torus%
\mathcal{)}$\ representations and then the $\mathbb{Z}_{2}$ symmetry is
manifested through the following splitting .
\begin{equation*}
\mathcal{H}^{o}=\mathcal{H}\oplus \overline{\mathcal{H}}.
\end{equation*}
where $\overline{\mathcal{H}}$ is the symmetric of $\mathcal{H}$. Note that
that the above decomposition has in fact a larger automorphism group
containing the orbifold discrete $\mathbb{Z}_{2}$ group just as a
subsymmetry. To
make an idea on this equivalence, it is enough to note that under the non
commutative orbifold coordinate choice eqs(\ref{orbicons}), one sees that $\mathcal{A}%
^{o}_{\theta }(\mathcal{O}rbifold)$ is simply
\begin{equation*}
\mathcal{A}^{o}_{\theta }(\mathcal{O}rbifold)=\mathcal{A}_{\theta }(Torus)\oplus
\overline{\mathcal{A}_{\theta }}(Torus)
\end{equation*}
where $\overline{\mathcal{A}_{\theta }}\mathcal{(}Torus\mathcal{)}$ is the
image of $\mathcal{A}_{\theta }\mathcal{(}Torus\mathcal{)}$ under $\mathbb{Z}_{2}.$ So operators of $%
\mathcal{A}^{o}_{\theta }\mathcal{(O}rbifold\mathcal{)}$ and too
particularly the projectors split into two irreducible factors
describing branes and anti branes.

\section{Conclusion}

In this paper we have studied tachyon condensation in non
commutative orbifolds. We have studied, amongst others, soliton
solutions for these compact manifolds and derived in particular
the generalization of the Power-Rieffels configurations on the
irrational non commutative orbifolds with a Z$_{2}$ discrete
symmetry. More precisely starting from an original non-BPS $D2l$
brane of mass M$_{D2l},l=1,2,...$ wrapped on
$\mathbb{T}^{2l}/\mathbb{Z}_{2}$, we have studied its condensation
using GMS formalism and related ideas and shown: (a) The existence
of general bound states extending both those
studied in \cite{g} and \cite{j} and containing amongst others $\overline{D}$-$\overline{%
D}$ bound states. (b) Using general relations on the brane
spectrum and T-duality, we have shown that the suggestion of Bars
{\it et al} regarding the interpretation of the $D0$-$D2$ branes,
which by the way extends to more general $D$-$D$ bounds, applies
as well to $\overline{D}0$--$\overline{D}2$ bound states and more
generally to $\overline{D}$--$\overline{D}$ bounds. (c) The
condensation incorporates naturally $D$--$\overline{D}$ brane
systems on quantum torii without needing to introduce an original
anti-$D$ brane. We have demonstrated the equivalence of the
$D$--$\overline{D}$ brane system we have obtained, the
$\mathbb{Z}_2$ symmetry of the orbifold do the full job. Finally
we would like to note that we expect that our results extend as
well to a large variety of toric orbifolds with other discrete
symmetries.\\

\vskip 1cm
\centerline{\large\bf{Aknowledgements}}

E. M. Sahraoui would like to thank the Physics department of the
university of Turin  for kind hospitality. He thanks also M. Frau,
P. Fre, I. Pesando and M. Billo for their helpful discussions.\\
He is grateful to M. Hssaini and A. Rhalami too for fruitful
comments.\\
 This work has been partially supported by SARS,''Programme
de Soutien \`{a} la Recherche Scientifique de l'universit\'{e}
Mohammed V-Agdal, Rabat''.\\


\newpage



\begin{thebibliography}{10}
\bibitem[1]{a}  A. Sen, ``Tachyon condensation on the brane anti-brane
system,'' \textbf{JHEP 9808:012}(1998), hep-th/9805170.

\bibitem[2]{b}  A. Sen, ``SO(32) spinors of type I and other solitons on
brane anti-brane pair,'' \textbf{JHEP 9809:023}(1998), hep-th/9808141.

\bibitem[3]{c}  J. A. Harvey, P. Kraus and F. Larsen, ``Exact
noncommutative solitons,'' hep-th/0010060.

\bibitem[4]{d}  D. V. Gross and N. A. Nekrasov, ``Monopoles and strings in
noncommutative gauge theory,'' hep-th/0005204.

\bibitem[5]{e}  R. Gopakumar, S. Minwalla and A.
Strominger,``Noncommutative solitons,'' hep-th/0003160.

\bibitem[6]{f}   J. A. Harvey, P. Kraus, F. Larsen and E. J. Martinec,
``D-branes and strings as noncommutative solitons,'' hep-th/0005031.


\bibitem[7]{g}  I. Bars, H. Kajiura, Y. Matsuo and T. Takayanagi, ``Tachyon
condensation on non-commutative torus,'' hep-th/0010101.

\bibitem[8]{h} Y. Matsuo, ``Topological charges of noncommutative soliton,'' hep-th/0009002.

\bibitem[9]{i}  J. A. Harvey and G. Moore, ``Noncommutative tachyons and K-theory,'' hep-th/0009030.

\bibitem[10]{j} E. M. Sahraoui and E. H. Saidi, ``Solitons on compact and noncompact spaces in large noncommutativity,'' hep-th/0012259.

\bibitem[11]{k} M. R. Gaberdiel, ``Discrete torsion orbifolds and D-branes,'' {\bf JHEP 0011:026}(2000), hep-th/0008230

\bibitem[12]{l} Emil J. Martinec and Gregory Moore,
"Noncommutative Solitons on Orbifolds", hep-th/0101199

\bibitem[13]{m} Tadashi Takayanagi, "Holomorphic Tachyons and Fractional
D-branes", hep-th/0103021

\bibitem[14]{n} T. Krajewski and M. Schnabl, "Exact solitons on
noncommutative tori", hep-th/0104090

\bibitem[15]{o} H. Kajiura, Y. Matsuo and T. Takayanagi, "Exact
tachyon condensation noncommutative torus", hep-th/0104143

\end{thebibliography}
\end{document}